\documentclass[aps,prl,twocolumn,showpacs,floatfix]{revtex4}
\usepackage{hyperref}
\usepackage{amsmath}
\usepackage{graphicx,epsfig,psfrag}
\usepackage{amssymb}
\usepackage{color}
\begin{document}
\title{Charge Kondo anomalies in PbTe doped with Tl impurities}
\author{T. A. Costi$^{1}$}
\author{V. Zlati\'c$^{1,2,3}$}
\affiliation{
$^{1}$Peter Gr\"{u}nberg Institut and Institute for Advanced Simulation, 
Forschungszentrum J\"{u}lich, D-52425 J\"ulich, Germany\\
$^{2}$Institute of Physics, HR-10001 Zagreb, Croatia\\
$^{3}$J. Stefan Institute, SI-1000 Ljubljana, Slovenia}
\begin{abstract}
We investigate the properties of PbTe doped with a small concentration 
$x$ of Tl impurities acting as acceptors and described by Anderson impurities 
with negative onsite correlation energy. We use the numerical 
renormalization group method to show that the resulting charge Kondo effect 
naturally accounts for the unusual low temperature and doping dependence
of normal state properties, including the self-compensation effect in the 
carrier density and the nonmagnetic Kondo anomaly in the resistivity. 
These are found to be in good qualitative agreement with experiment. Our
results for the Tl $s$-electron spectral function provide a new interpretation 
of point contact data. 
\end{abstract}
\vspace{-4.8cm}
%
%
\date{\today}
\maketitle
{\em Introduction.---}
PbTe is a narrow gap semiconductor with a band gap of $190$\,{\rm meV} at zero
temperature \cite{ravich.98}. Upon doping with Tl impurities \cite{ravich.98,volkov.02} 
a number of striking anomalies appear in the low temperature properties. 
These anomalies, described below, are currently not well understood. 
They are believed to be due to the special nature of Tl impurities 
in PbTe, which have been proposed to act as negative $U$ centers, 
where $U<0$ is an attractive onsite Coulomb energy. In compounds, the 
outer $6$\,{\rm s} shell of Tl can be empty, singly occupied or doubly occupied.
A negative Coulomb correlation results in the nonmagnetic empty and doubly 
occupied states being lower in energy than the magnetic singly occupied state. 
Hybridization of the {\rm s}-states with the valence band states of PbTe can then 
result, via dynamic valence fluctuations, in a (nonmagnetic) charge Kondo (CK) 
analogue \cite{taraphder.91,andergassen.11} of the conventional spin 
Kondo effect \cite{hewson.97}. This nonmagnetic CK 
effect has been argued to explain many of the observed properties of 
Pb$_{1-x}$Tl$_{x}$Te \cite{matsushita.05,dzero.05,matsushita.06,matusiak.09,erickson.10}
which would then constitute the first physical realization of this effect.

A remarkable feature of Pb$_{1-x}$Tl$_{x}$Te is that the low temperature properties 
depend sensitively on the Tl concentration, with a qualitatively 
different behavior below and above a critical concentration $x_c\simeq 0.3$ 
\,\% Tl. For example, while Pb$_{1-x}$Tl$_{x}$Te remains metalliclike
down to the lowest temperatures for $x<x_{c}$, for $x>x_{c}$ it 
becomes superconducting with a transition temperature T$_c(x)$ increasing 
linearly with $x$ and reaching $1.5$\, {\rm K} at $x=1.5$ \,\% Tl 
\cite{chernik.81,matsushita.05}. This is surprisingly high given the low 
carrier density of less than 10$^{20}$ holes/{\rm cm}$^3$. In addition,
measurements of the Hall number $p_H=1/R_H e$ 
\cite{murakami.96,matsushita.06} indicate that the number of holes grows linearly with 
$x$ for $x < x_c$, whereas for $x>x_{c}$ the number of holes remains almost 
constant: the system exhibits ``self-compensation'' and the chemical 
potential is pinned to a value $\mu=\mu^{*}\simeq 220$\,{\rm meV}
\cite{murakami.96,kaidanov.89}. 
Transport measurements also show anomalous behavior at low temperatures: 
while for $x<x_{c}$, the residual resistivity, $\rho_0$, is very small 
and almost constant as a function of $x$, for $x>x_{c}$, 
$\rho_{0}$ increases linearly with $x$ \cite{matusiak.09}. 
For $x<x_{c}$, the resistivity, $\rho(T)$, exhibits a positive slope at low temperature, 
while for $x>x_{c}$, the slope is negative and a Kondo-like  
contribution, $\rho_{imp}(T)$, is observed for $T\lesssim 10K$ 
\cite{matsushita.05,matusiak.09}. 
The origin of this anomaly is not due to magnetic impurities, since
the susceptibility is diamagnetic \cite{matsushita.05}. 

In this Letter, we show that the unusual concentration and 
low temperature dependence of a number of normal state properties of 
Pb$_{1-x}$Tl$_{x}$Te can be naturally explained within a picture of 
dilute Tl impurities acting as negative $U$ centers in the PbTe host. 
Such a dilute impurity description is suggested by the observed 
concentration dependence of many properties, as summarized above, 
e.g., the linear dependence on $x$ of the residual resistivity for 
$x>x_{c}$. We model the Tl impurities by the negative $U$ 
Anderson model \cite{anderson.75} and solve this for the static and dynamic
properties by using the numerical renormalization group (NRG) method
\cite{nrg}. Our results, with comparisons to experimental data,  
and previous experimental and theoretical work on Pb$_{1-x}$Tl$_{x}$Te 
\cite{matsushita.05,dzero.05,malshukov.91}, provide strong evidence that 
the CK effect \cite{taraphder.91} is realized in this system. 

The idea that doping PbTe with Tl induces resonant states in 
the valence band of PbTe was conjectured early on (see 
Ref.~\onlinecite{ravich.98}). These states are mainly of Tl {\it s}-character 
and lie close to the top of the valence band, as recently shown
by density functional theory calculations \cite{xiong.10}. 
Resonant states alone, however, and generalizations of this to an impurity
band of resonant states, cannot explain properties such as the  
superconductivity of Pb$_{1-x}$Tl$_{x}$Te. For this, a coupling of Tl ions to
the lattice \cite{shelankov.87}, or a static mixed valence model \cite{drabkin.81} have
been proposed.
In the latter, Tl impurities, known to be valence skippers in compounds, are 
assumed to dissociate into energetically close Tl$^{1+}$ (6{\it s}$^2${\it p}$^1$) 
and Tl$^{3+}$ (6{\it s}$^0${\it p}$^3$) ions,
while the Tl$^{2+}$ (6{\it s}$^1${\it p}$^2$) configuration lies higher
in energy \cite{weiser.79}. In the highly polarizable PbTe host, this  
can result in negative on-site $U$ and provides a mechanism for superconductivity 
and a qualitative explanation for the observed self-compensation, 
chemical potential pinning and the diamagnetic behavior 
of Pb$_{1-x}$Tl$_{x}$Te \cite{drabkin.81}. For a more quantitative
explanation of the observed anomalies, and in order to explain the
Kondo anomalies in the resistivity, a more realistic model is needed, 
which includes dynamic fluctuations between the Tl$^{1+}$ and Tl$^{3+}$ valence 
states. This motivates our use of the negative $U$ Anderson model 
\cite{anderson.75}, as formulated for Pb$_{1-x}$Tl$_{x}$Te in Ref.~\onlinecite{dzero.05} 
and discussed as a model for this system in 
Refs.~\onlinecite{malshukov.91,dzero.05,matsushita.05,matsushita.06,matusiak.09,
erickson.10,nakayama.08}.
For completeness, we mention also the valence band model for
PbTe, in which the main effect of Tl doping is assumed to be a rigid shift 
of the chemical potential into the valence band. This model may be relevant 
for the transport properties of PbTe doped with Tl impurities \cite{singh.10}
at high temperature ($T> 300 K$), where the charge Kondo effect 
is suppressed. It fails, however, to describe the low temperature anomalies 
that we are addressing in this Letter, e.g. the Kondo upturn in the 
resistivity at $T<10 K$ for $x>x_{c}$.

{\em Model and calculations.---}
We consider $n$ Tl impurities in a PbTe crystal with $N$ Pb sites
described by the Hamiltonian $H=H_{band} +H_{imp}+H_{hyb}$. The 
first term, $H_{band} =  \sum_{{\bf k}\sigma} (\epsilon_{\bf k} -\mu_{e}) 
c^{\dagger}_{ {\bf k}\sigma}c_{ {\bf k}\sigma}$, describes the valence $p$-band 
of PbTe, where $\mu_{e}$ is the (electron) chemical potential and $c^{\dagger}_{{\bf k}\sigma} $ 
creates an electron with energy $\epsilon_{\bf k}$. The second term,
$H_{imp}=  (\epsilon_0 -\mu_{e}) \sum_{i=1\sigma}^{n} 
\hat n_{is\sigma} + U \sum_{i=1}^{n} n_{is\uparrow} n_{is\downarrow}$,
describes the Tl impurities, where 
$n_{is\sigma} =s^{\dagger}_{i\sigma} s^{}_{i\sigma}$ is the number operator for 
a Tl {\it s}-electron at site $i$ with spin $\sigma$ and energy $\epsilon_0$, and  
$U$ is the (negative) correlation energy. The last term,
$H_{hyb} = \sum_{i=1}^{n}\sum_{k\sigma}V_{0}
(c_{{\bf k}\sigma}^{\dagger}s_{i\sigma} + h.c.)$, models the hybridization of
Tl $s$-states with the valence band $p$-states and $V_{0}$ is the matrix element for
the {\it s}-{\it p} interaction. Its strength is characterized by the hybridization function 
$\Delta(\omega)=\pi V_{0}^{2}\sum_{\bf k}\delta(\omega-\epsilon_{{\bf k}\sigma})=\pi V_{0}^{2}
{\cal N}(\omega)$, where we retain the full energy dependence of the $p$-band density of states 
${\cal N}(\omega)=\sum_{\bf k}\delta(\omega-\epsilon_{\bf k})$ of PbTe \cite{singh.10}.

The chemical potential $\mu_{e}$ determines $n_e=\frac{1}{N}\sum_{{\bf k}\sigma}
\langle c^{\dagger}_{ {\bf k}\sigma}c_{ {\bf k}\sigma}\rangle
$ and $n_s=\frac{1}{n}\sum_{i=1}^{n}\sum_{\sigma}\langle n_{is\sigma}\rangle$,  
the average number of {\it p} and {\it s} electrons per site. We denote by
$x=n/N$ the concentration of Tl impurities. Since Tl acts as an acceptor, the ground state 
corresponds to the Tl$^{1+}$ ($n_s=2$) configuration and the Tl$^{3+}$ ($n_s=0$) 
configuration is split-off from the ground state by the energy 
$\delta =E({\rm Tl}^{3+})-E({\rm Tl}^{1+})>0$. A concentration 
$x$ of Tl impurities accommodates $x(n_{s}-1)$ electrons (per Tl site), where the number of accepted 
electrons in the 6{\it s} level of Tl is measured relative to the neutral Tl$^{2+}$ ({\it s}$^{1}$) 
configuration having $n_{s}=1$. These electrons are removed from the valence band leaving
behind $n_{0}=1-n_{e}$ holes. Thus, charge neutrality implies $n_{0}=x(n_{s}-1)$\, \cite{dzero.05},
which for a given $x$ and temperature $T$ has to be satisfied by adjusting the chemical potential 
$\mu_{e}$. Here, we neglect interimpurity interactions and solve $H$ for a 
collection of single independent negative-$U$ centers by using the 
NRG \cite{nrg}. For each $x$ and each $T$ we satisfy 
the above equation by self-consistently determining the chemical potential $\mu_{e}$ 
(or equivalently the hole chemical potential $\mu$, which we henceforth use). 

The electrical resistivity of electrons scattering from a dilute concentration $x$ of Tl 
impurities is obtained from the 
usual expression, ${\rho_{imp}}(T)=1/e^2 L_{11}$, where $ L_{11}$ is the static limit of the 
current-current correlation 
function. In the absence of nonresonant scattering the vertex corrections
vanish and the relevant transport integral $L_{11}$ 
can be written as \cite{costi.94}, 
\begin{equation}
\label{eq: lij_final}
L_{11} = \sigma_{0}\int_{-\infty}^{\infty}d\omega
\left(
-\frac{\partial f(\omega)}{\partial\omega} \right){\cal N}(\omega)\tau(\omega,T),
\end{equation}
where $\sigma_{0}=\langle v_{k_{F}}^{2}\rangle$ is the velocity factor $v_{k}^{2}$ 
averaged over the Fermi surface, $f(\omega)=1/[1+\exp(\omega/k_B T)]$ is the Fermi function,
$\tau(\omega,T)$ is the conduction-electron transport time\cite{costi.94} 
${\tau(\omega,T)}^{-1} = 2n_{\rm Tl}  V_{0}^2 A(\omega,T)$
and $A(\omega,T)=-\frac{1}{\pi}\mbox{Im}\; G(\omega + i 0^{+})$ is
the  spectral function of {\it s\/}-electrons. The number of Tl impurities 
$n_{\rm Tl}$ per {\rm cm}$^3$ is related to $x$ in percent by 
$n_{\rm Tl}=1.48\times 10^{20}x$ \cite{note-conversion}.

{\em Choice of model parameters.---}
We use $\mu^{*}\approx 225$\,{\rm meV}, close to the value obtained from
tunneling experiments \cite{murakami.96,kaidanov.89}.
Other parameters, such as $\Delta_{0}=\Delta(\mu^{*})$, required to fix the hybridization
function $\Delta(\omega)$, and $U$ are largely unknown. Our interpretation of the tunneling
spectra (see discussion of Fig.~\ref{fig2} below), suggests $U\approx -30\, {\rm meV}$.
The measured Kondo-like resistivity for $x\ge x_{c}$ at low temperatures 
requires that $|U|\gg \Delta_{0}$. We take $U/\Delta_{0}=-11$ with 
$\Delta_{0}=2.7\, {\rm meV}$ to yield a Kondo temperature  $T_{K}\approx 1.23\, {\rm K}$ 
below the highest $T_{c}\approx 1.5\,{\rm K}$ at 
$x=1.5 \,\% \gg x_{c}$, where $T_{K}$ is defined via the impurity resistivity 
$\rho_{imp}(T=T_{K})=0.5\rho_{imp}(T=0)$ for $x\gg x_{c}$. 
The overall qualitative aspects of our results remained the 
same for values of $U$ in the range $10$\,{\rm meV} $\leq |U| \leq 220$\,{\rm meV} 
and $|U|/\Delta_{0}\gg 1$.

\begin{figure}[t]
\includegraphics[width=\linewidth,clip]{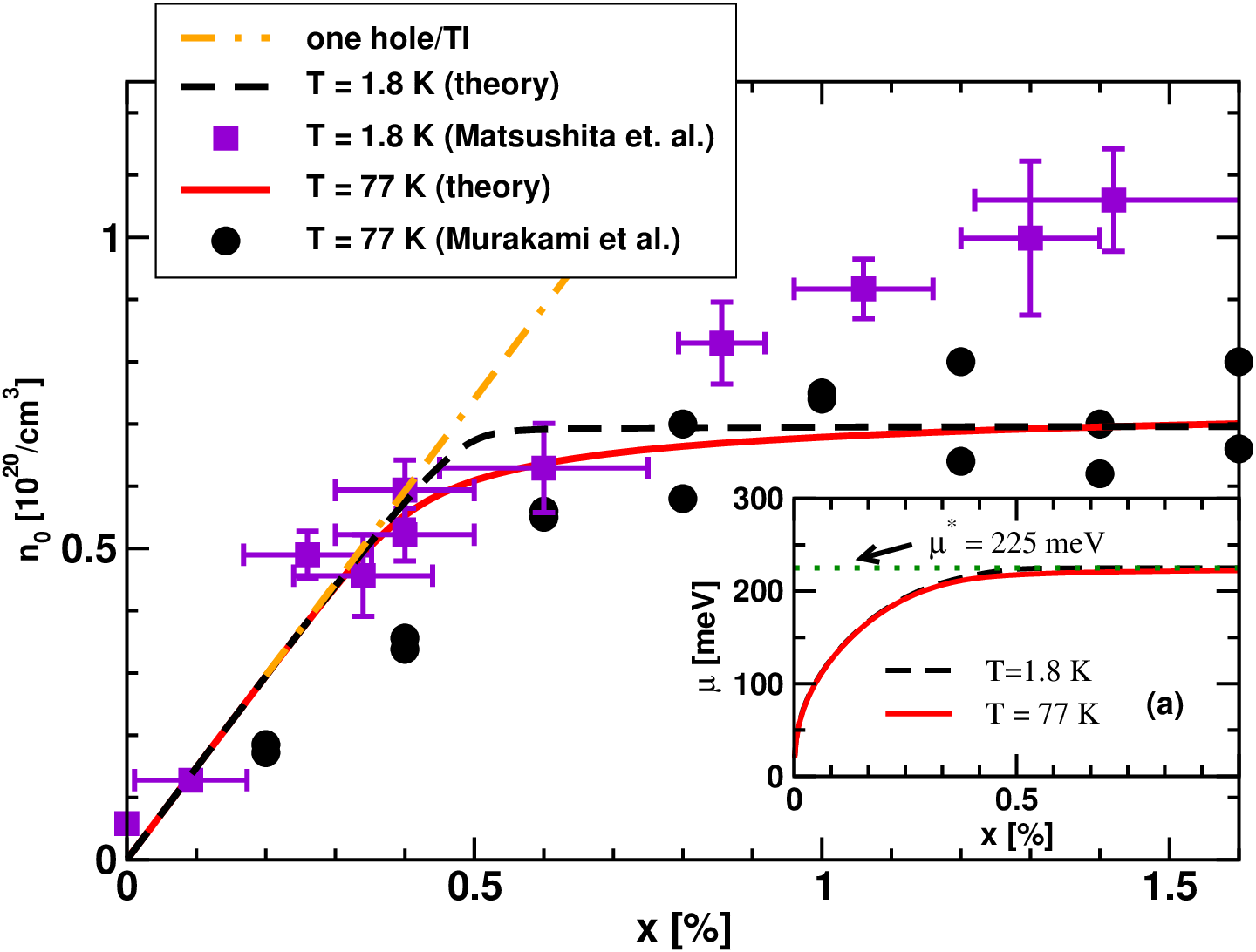}
\caption{{\em (Color online)} Hole carrier density $n_{0}$ versus Tl doping $x$ in \,\% for 
$T=1.8\, {\rm K}$ and $T=77$\, {\rm K}. Filled circles: 
experimental data at $T=77$\, {\rm K} \cite{murakami.96}, filled squares: experimental 
data at $T=1.8\,{\rm K}$ \cite{matsushita.06}, dot-dashed line: expected $n_{0}$ for one hole per Tl. 
Inset (a): hole chemical potential $\mu$ versus $x$ at $T=1.8\, {\rm K}$ and $T=77\, {\rm K}$ [and hence
$\delta(\mu)=2(\mu-\mu^{*})$)].
\label{fig1}}
\end{figure}
{\em Qualitative considerations.---}
It is instructive to first make some qualitative remarks, starting
from the atomic limit $V_{0}=0$ \cite{dzero.05}. 
For $x=0$ the chemical potential lies in the gap between the valence and conduction bands.
For finite but very small $x$ each Tl impurity accepts one electron, 
i.e. $n_{s}\approx 2$ and $n_{0}=x(n_{s}-1)\approx x$ 
grows linearly with $x$. At the same time, the 
chemical potential shifts downwards into the valence band $\mu < E_{v}$, where $E_{v}$ denotes
the top of the valence band. This implies that the splitting $\delta(\mu)=-(2(\epsilon_0-\mu) + U)$ 
between donor and acceptor configurations decreases. Eventually, at a critical concentration
$x=x^{*}$, the chemical potential reaches $\mu=\mu^{*}=\varepsilon_{0}+U/2$ where 
$\delta(\mu)=2(\mu-\mu^{*})=0$ and the system is in a (static) mixed valence state where 
the Tl$^{1+}$ and Tl$^{3+}$ configurations are degenerate. In this situation
$n_{s}=1$, and further doping cannot increase the hole carrier density beyond
the value $n_{0}(\mu^{*})$, i.e. one has self-compensation with a pinning 
of the chemical potential to $\mu^{*}$ \cite{dzero.05}. 

For finite $V_{0}$, quantum fluctuations between the degenerate states Tl$^{1+}$ and Tl$^{3+}$ at
$\mu=\mu^{*}$ become important and lead to a CK effect. This significantly affects
all static and dynamic properties and needs to be taken into account in describing the
experiments. It is also important for $\mu>\mu^{*}$, since 
a finite charge splitting $\delta(\mu)> 0 $ in the negative-$U$ Anderson model 
is similar to a Zeeman splitting in the conventional spin Kondo effect \cite{iche.72}. 
The latter is known to drastically influence all properties \cite{hewson.97}. 
Thus, for the whole range of concentrations $x$, one expects fluctuations 
to play an important role in the properties of Pb$_{1-x}$Tl$_{x}$Te.

{\em Numerical results.---}
Figure~\ref{fig1} shows $n_{0}(x)$ versus $x$ at $T=1.8\,{\rm K}$ and at 
$T=77\,{\rm K}$ and a comparison with experimental data on Hall number 
measurements \cite{murakami.96,matsushita.06}. At low dopings, $n_{0}$ is 
linear in $x$ both in theory and in experiment, as expected for Tl impurities
acting as acceptors (dot-dashed curve in Fig.~\ref{fig1}). 
However, the efficiency, $n_{0}(x)/n_{\rm Tl}(x)$, of Tl dopants 
in supplying holes at low $x$ is only around $65$ \% in the data of 
Ref.~\onlinecite{murakami.96} as opposed to $100$\,\% in our model 
calculations and in the data of Ref.~\onlinecite{matsushita.06}.
At higher dopings $n_{0}(x)$ saturates rapidly  with increasing $x$ 
for $T=1.8\,{\rm K}$ and more slowly at higher temperatures. 
The theoretical crossover from linear to saturated behavior occurs at 
$x=x^{*}\approx 0.5\,\%$, larger than the value $x_{c}\approx 0.3\,\%$ 
for the onset of superconductivity. The theoretical saturation density 
$n_{0}\approx 0.7\times 10^{20}/{\rm cm}^{3}$ is close to the experimental value \cite{murakami.96}.
The self-compensation effect for $x\gg x^{*}\approx 0.5\,\%$ is a characteristic signature of 
the CK state: on entering this state the Tl ions fluctuate 
between Tl$^{1+}$ and Tl$^{3+}$ so the average valence is Tl$^{2+}$,
which corresponds to no additional electrons being accepted or donated. 
The CK state may be destroyed by lifting the degeneracy of the pseudospin states, 
i.e. in the language of the spin Kondo effect, by applying a ``magnetic field''. 
A ``magnetic field'' in the CK effect corresponds to doping or shifting $\mu$. This
has been achieved by counterdoping with In ions \cite{erickson.10}, which act 
as donors. The Kondo anomalies, e.g. in the resistivity, were observed to
vanish, providing further support to the CK picture. The self-compensation
effect can be seen also in the pinning of the hole chemical potential $\mu$ for $x>x^{*}$, 
shown in Fig.~\ref{fig1}a. For $x<x^{*}$, $\mu$ grows nonlinearly with $x$ and 
rapidly approaches the value $\mu^{*}=225$\,{\rm meV} for $x>x^{*}$, both at $T=1.8\, {\rm K}$ 
and at $T=77\,{\rm K}$. 

\begin{figure}[t]
\includegraphics[width=\linewidth,clip]{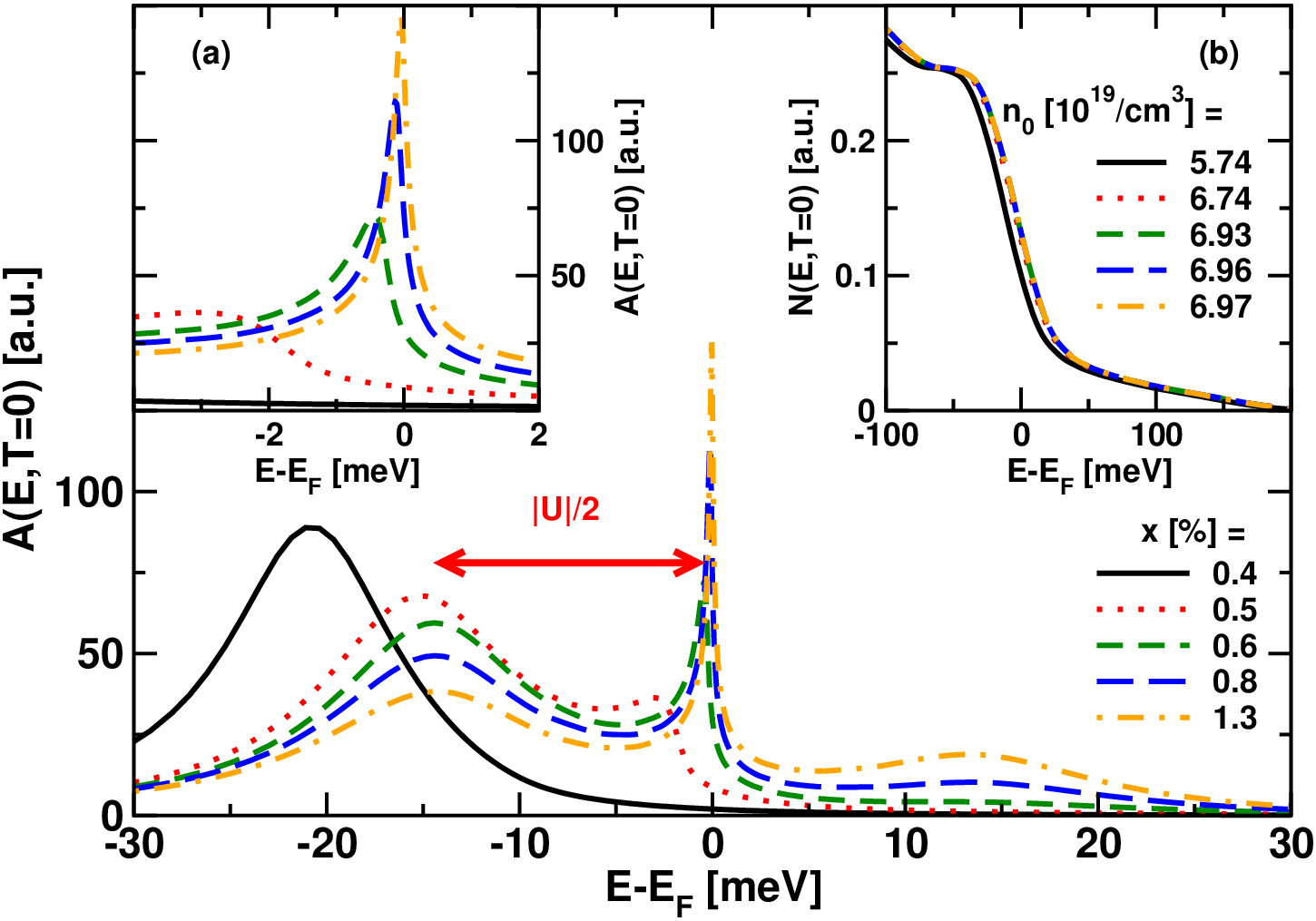}
\caption{{\em (Color online)} Tl spectral 
function $A(E,T=0)$ versus $E-E_{F}$ for a range of Tl dopings $x$. 
Inset (a): region near $E=E_{F}$ showing the
CK resonance. Inset (b): PbTe valence band density of states ${\cal N}(E)$ 
versus $E-E_{F}$ for $x$ as in the main panel \cite{singh.10} (legend: 
hole densities $n_{0}$ for each $x$).
\label{fig2}}
\end{figure}
Figure~\ref{fig2} shows the Tl {\it s\/}-electron 
spectral function $A(E,T=0)$ at zero temperature and different dopings $x$. 
For small doping, $x<x^{*}$, the hole chemical potential $\mu$ 
lies above $\mu^{*}$ within the shallow part of the valence band density of states ${\cal N}(E)$
(see Fig.~\ref{fig2}b), consequently the splitting $\delta(\mu) = 2(\mu-\mu^{*})$, see Fig.~\ref{fig1}a, 
is large. Such a
large splitting acts like a large Zeeman splitting in the conventional positive $U$ Anderson
model and polarizes the spectral function so that its weight lies mostly below the Fermi level
$E_{F}$ \cite{andergassen.11}. For $x\ge x^{*}$, $\mu$ approaches $\mu^{*}$ and a CK effect 
develops. The spectral function develops a sharp asymmetric Kondo resonance close to,
but below $E_{F}$ (see Fig.~\ref{fig2}a)  and an upper Hubbard satellite peak appears 
above $E_{F}$. Early tunneling experiments
for low dopings $x< 0.3\,\%$ showed only one resonant level below $E_{F}$ \cite{kaidanov.89}, 
whereas more recent tunneling
experiments for $x>0.6\,\%$ show two ``quasilocal'' peaks \cite{murakami.96}, a narrow one
of width $6\,{\rm meV}$ close to $E_{F}$ and a broader one of width $12\, {\rm meV}$ at a
nearly constant energy 13-15 {\rm meV} below this. Interpreting these as the Kondo resonance 
and the lower Hubbard satellite peaks in Fig.~\ref{fig2} yields $|U|/2\approx 15 \,{\rm meV}$ 
and hence the value $U\approx -30\,{\rm meV}$ used in our calculations. 
Since the CK resonance is temperature dependent \cite{andergassen.11}, 
the above interpretation could be tested with temperature dependent studies of tunneling 
or high resolution photoemission spectra.

\begin{figure}[t]
\includegraphics[width=\linewidth,clip]{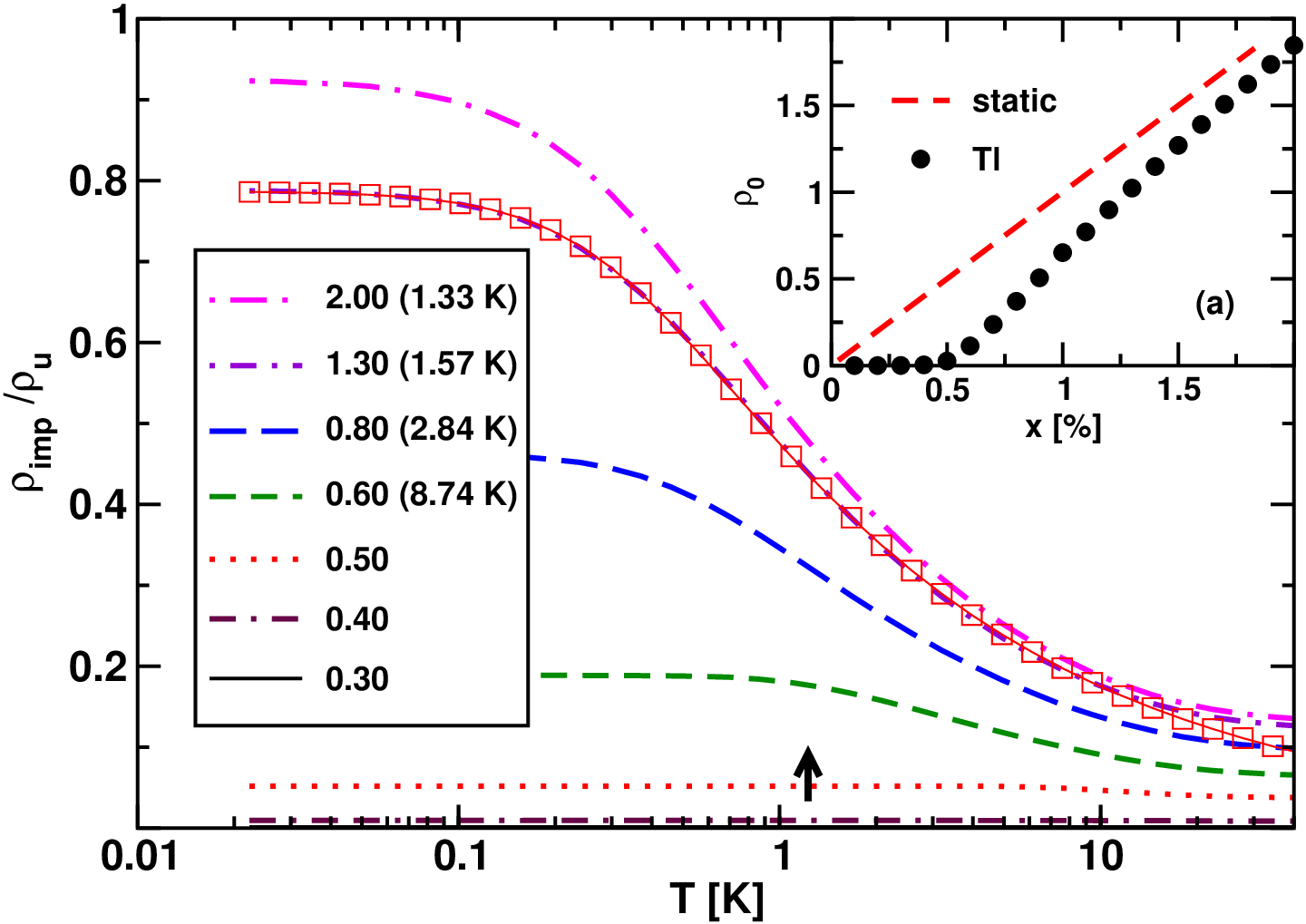}
\caption{{\em (Color online)} Impurity resistivity 
$\rho_{\rm imp}(T)/\rho_{u}$ versus temperature $T$ and
a range of Tl concentrations $x$ (in percent, legend: column 1). For
$x>x^{*}$, $\rho_{imp}$ is well described by the 
usual spin Kondo resistivity \cite{costi.94} (shown as open squares for $x=1.3\,\%$),  
but with effective Kondo scales $T_{K}^{eff}$ (legend: column 2).  
$\rho_{u}=2n_{\rm Tl}/(e^{2}\pi \hbar N_{F}^{2}\sigma_{0})$ 
is the residual resistivity for unitary scatterers
with $N_{F}={\cal N}(\mu^{*})$. The vertical arrow: $T_{\rm K}=1.23\,{\rm K}$. Inset (a): 
$\rho_{0}=x\rho_{\rm imp}(T=0)/\rho_{u}$ versus $x$ (filled circles). 
Dashed line: $\rho_{0}$ for static unitary scatterers in place of Tl.
\label{fig3}}
\end{figure}

The temperature and doping dependence of the impurity resistivity, $\rho_{imp}$,
is shown in Fig.~\ref{fig3}. 
For $x<x^{*}$, Tl impurities act as acceptors with a well-defined valence
state (Tl$^{1+}$). They therefore act as weak potential scatterers 
and consequently the resistivity is much below the unitary value, as
seen in Fig.~\ref{fig3}. For $x>x^{*}$,
dynamic fluctuations between the nearly degenerate Tl$^{1+}$ and Tl$^{3+}$ states
leads to the CK effect and $\rho_{imp}$ approaches the resistivity 
for unitary scatterers at $T=0$. For $x>x^{*}$, $\rho_{imp}$ is well described by the 
{\em spin} Kondo resistivity \cite{costi.94} (open squares, Fig.~\ref{fig3}) 
with a logarithmic form around $T\approx T_{K}^{eff}$, where $T_{K}^{eff}$ is an effective Kondo scale, 
and $T^{2}$ Fermi liquid corrections at low $T\ll T_{K}^{eff}$, in qualitative
agreement with experiment \cite{matsushita.05}. The effective 
Kondo scale $T_{K}^{eff}$ is a function of the charge splitting 
$\delta(\mu)$ and $T_{K}$, and approaches the true Kondo scale $T_{K}$
only asymptotically for $x\gg x^{*}$ (see legend to Fig.~\ref{fig3}). 
Finally, Fig.~\ref{fig3}a shows that the impurity residual resistivity is
significant only when the CK effect is operative, i.e. for $x>x^{*}$, in
qualitative agreement with experiment \cite{matsushita.05,matusiak.09}.

{\em Conclusions.---}
In summary, we investigated the normal state properties of 
Pb$_{1-x}$Tl$_{x}$Te within a model of Tl impurities acting as negative $U$ centers. 
Our NRG calculations support the suggestion that the CK effect is realized in 
Pb$_{1-x}$Tl$_{x}$Te \cite{matsushita.05,dzero.05}. They explain a number of low
temperature anomalies of Pb$_{1-x}$Tl$_{x}$Te, including the qualitatively 
different behavior below and above the critical concentration $x^{*}$, 
where $x^{*}\approx 0.5\,\%$ is close to $x_{c}\approx 0.3\,\%$ for the onset of
superconductivity.  At $x=x^{*}$, two 
nonmagnetic valence states of Tl become almost degenerate and the ensuing pseudospin 
CK effect results in a Kondo anomaly in the resistivity for $x>x^{*}$ and a 
residual resistivity approximately linear in $x$.
Our results for these quantities and the carrier density $n_{0}(x)$
are in good qualitative agreement with experiments 
\cite{matsushita.05,matsushita.06,matusiak.09,murakami.96}. 
For the Tl $s$-electron spectral function, we predict that one peak should be 
present far below $E_{F}$ for $x<x^{*}$ and that a second temperature dependent 
Kondo resonance peak develops close to, but below $E_{F}$, 
on increasing $x$ above $x^{*}$. This provides a new interpretation of measured 
tunneling spectra\cite{murakami.96}, which could be tested by temperature 
dependent studies of tunneling or photoemission spectra. In the future, it would
be interesting to extend this work to include the effects of disorder 
within an Anderson-Hubbard model description.
\acknowledgments We thank K. M. Seemann, D. J. Singh, H. Murakami, P. Coleman, 
G. Kotliar, and I. R. Fisher for discussions and 
D. J. Singh, H. Murakami and I. R. Fisher for data \cite{singh.10,matsushita.06}. 
V.Z. acknowledges support by Croatian MZOS Grant No.0035-0352843-2849, 
NSF Grant DMR-1006605 and Forschungszentrum J\"{u}lich.
T. A. C. acknowledges supercomputer support from the John von Neumann Institute for 
Computing (J\"{u}lich).

\end{document}